\documentclass[aps,prd,eqsecnum,preprint]{revtex4}
\usepackage{color,graphicx}
\usepackage{amsfonts}
\usepackage{amssymb}
\usepackage{bm}

\newcommand{\brho}{\ensuremath{\bar{\rho}}}
\newcommand{\ka}{\ensuremath{\kappa }}
\newcommand{\ep}{\ensuremath{\epsilon }}

\begin{document}

{\baselineskip0pt
\rightline{\large\baselineskip16pt\rm\vbox to20pt{\hbox{OCU-PHYS-482}
            \hbox{AP-GR-147}
\vss}}%
}


\title{The gravitational wave background induced by nonlinear interactions between isotropic inhomogeneities and anisotropic inhomogeneous density perturbations}
\author{
Hiroyuki Negishi\footnote{Electronic address:negishi@sci.osaka-cu.ac.jp}
}

\affiliation{
Advanced Mathematical Institute,
 Osaka City University, 
Osaka 558-8585, Japan
}

\begin{abstract}
Usually, we assume that there is no inhomogeneity isotropic in terms of our location in the universe.
This assumption has not been observationally confirmed yet in sufficient accuracy and we need to consider a method to restrict isotropic inhomogeneities more strongly.
If there are isotropic inhomogeneities in the universe, the gravitational wave background is induced by nonlinear interactions between isotropic inhomogeneities and anisotropic inhomogeneous density perturbations.
In this paper, we calculate the power spectrum and the relative energy density of this gravitational wave background to discuss observability.
We show that, at the decoupling time, the relative energy density of this gravitational wave background is comparable to that of the primordial inflationary gravitational wave background whose tensor-to-scalar ratio is 0.001, if  $6\%$-level isotropic deviations from homogeneity exist.
Therefore, there are possibilities of observing this gravitational wave background and restricting isotropic inhomogeneities by using  the future cosmic microwave background observation project.
\end{abstract}

\maketitle

\vskip1cm

\section{Introduction}\label{Sec1}
In the standard model of cosmology, we assume that the universe is isotropic and homogeneous on large scales. 
It is important that we observationally confirm this assumption to construct precision cosmology.
Isotropy whose symmetry center coincide with our location of the universe can be observationally confirmed by using observations in various directions and observed isotropy of the cosmic microwave background (CMB) with high accuracy of about $10^{-5}$ implies isotropy of the universe.
Verification of homogeneity of the universe is more difficult than verification of isotropy.
This is because our observations are confined on a past light cone and observational data includes information on the temporal evolution and spatial inhomogeneities of the universe.
In a recent study, isotropic inhomogeneities whose symmetry center coincide with our location have been restricted with combined observables and $10\%$-level deviations from homogeneity is permitted\cite{Redlich:2014gga}.
If there are isotropic inhomogeneities in the universe and we interpreted observational data under the assumption that the universe is homogeneous and isotropic on large scales, systematic errors on observational results occur.
Systematic errors on the amount of dark energy due to $10\%$-level deviations from homogeneity is comparable to the error caused in observation \cite{Negishi:2015oga}.
Thus, $10\%$-level deviations from homogeneity is not small and more stronger restriction on isotropic inhomogeneities is required to construct precision cosmology.

In previous studies of restricting isotropic inhomogeneities, the influence of isotropic inhomogeneities on anisotropic inhomogeneous perturbations was not taken into much consideration.
This is because theoretical prediction of this influence is difficult and the influence on observables was considered small.
In recent years, this influence becomes important, since the accuracy of observation is improving.
One of this influence is that gravitational potentials produced by isotropic inhomogeneities affect distributions of anisotropic inhomogeneous density perturbations and it is investigated by some authors\cite{Clarkson:2009sc,Zibin:2008vj,Nishikawa:2012we,Nishikawa:2013rna,Nishikawa:2014sga,February:2013qza,Meyer:2014qla}.
Another one of this influence is that nonlinear interactions between isotropic inhomogeneities and anisotropic inhomogeneous density perturbations induce the gravitational wave background.
This gravitational wave background has not been investigated so far and we investigated it in this paper.
The gravitational wave background is important perturbation variables to restrict isotropic inhomogeneities, since the gravitational wave background has a great influence on the CMB polarization and it become possible to observe the CMB polarization with high accuracy.
We calculate the relative energy density of the gravitational wave background induced by nonlinear interactions between isotropic inhomogeneities and anisotropic inhomogeneous density perturbations and compare it with that of the primordial inflationary gravitational wave background to discuss observability.

In this paper, to analyze the gravitational wave background induced by nonlinear interactions between isotropic inhomogeneities and anisotropic inhomogeneous density perturbations, we refer to the method that Nishikawa et al.\cite{Nishikawa:2012we} have used to analyze anisotropic inhomogeneous density perturbations affected by isotropic inhomogeneities.
We assume that there are isotropic inhomogeneous density perturbations with small amplitude in the universe and, on large scales, we describe the universe as the Friedmann-Lema\^{\i}tre-Robertson-Walker (FLRW) universe model with isotropic inhomogeneous perturbation.
In order to calculate the gravitational wave background induced by nonlinear interactions between isotropic inhomogeneities and anisotropic inhomogeneous density perturbations, we add the anisotropic inhomogeneous perturbation to this universe model and solve perturbation equations up to the order that isotropic inhomogeneities and anisotropic inhomogeneous density perturbations are coupled.

The organization of this paper is as follows.
In Sec.~II, we review  isotropic inhomogeneities. 
In Sec.~III, we derive evolution equations of the gravitational wave background induced by nonlinear interactions between isotropic inhomogeneities and anisotropic inhomogeneous density perturbations.
In Sec.~IV, we show the numerical result and discuss observability.
Finally, Sec.~V is devoted to the summary and discussion.

In this paper, we adopt following conventions:
Greek indices, $\mu, \nu$, run over the four spacetime coordinate labels.
Latin indices, $i, j$, and so on, run over the three spatial coordinate.
The geometrized unit in which the speed of light and Newton’s gravitational constant are one.

\section{Isotropic inhomogeneities}\label{Sec2}

As mentioned in introduction, we describe the universe as the FLRW universe model with isotropic inhomogeneous perturbation on large scales. 
Linear perturbations in the FLRW universe model have gauge freedom.
In this paper, we use Newtonian gauge.
By adopting Newtonian gauge, the infinitesimal world interval of the FLRW universe model with isotropic inhomogeneous perturbation is written in the form, 
\begin{eqnarray}
ds^2&=&a^2(\eta)\left[ -\left(1+2\ka \phi ^{(\ka)}(\eta ,r) \right)d\eta^2 +\left(1-2\ka \psi ^{(\ka)}(\eta ,r)\right)
\left( dr^2+r^2d\Omega ^2\right) \right],
\label{metric}
\end{eqnarray}
where $\ka$ is a positive dimensionless small parameter ($0<\ka \ll 1$), $a(\eta)$ is the scale factor scaled so as to be unity at the present time $\eta=\eta_0$, $\phi ^{(\ka)}$ and $\psi ^{(\ka)}$ are isotropic inhomogeneous perturbation and $d \Omega^2$ is the line element of the unit 2-sphere.

We assume that our universe model is filled with only the non-relativistic matter and the cosmological constant $\Lambda$.
The background  stress-energy tensor of the non-relativistic matter $\bar{T}_{\mu\nu}$ is given by
\begin{eqnarray}
\bar{T}_{\mu\nu}(\eta )=\bar{\rho}(\eta) \bar{u}_\mu (\eta)\bar{u}_\nu (\eta),
\label{T_ab} 
\end{eqnarray}
and the isotropic inhomogeneous  perturbation of the stress-energy tensor of the non-relativistic matter $T_{\mu\nu}^{(\ka)}$ is given by
\begin{eqnarray}
T_{\mu\nu}^{(\ka)} (\eta ,r)= \ka \bar{\rho}(\eta) \left(  \delta ^{(\ka)} (\eta,r) \bar{u}_\mu (\eta)\bar{u}_\nu (\eta)+ \bar{u}_\mu (\eta)  u_{\nu}^{(\ka)}(\eta ,r) + u_{\mu}^{(\ka)}(\eta,r)  \bar{u}_\nu (\eta) \right),
\label{Tka_ab} 
\end{eqnarray}
where $\bar{\rho}$ and $\delta ^{(\ka)}$ are the background energy density and the isotropic inhomogeneous  density perturbation respectively,  $\bar{u}_\mu$ and  $u_{\mu}^{(\ka)}$ are the background 4-velocity and the isotropic inhomogeneous  perturbation of the 4-velocity respectively.
The coordinate system is chosen so that the components of the 4-velocity are given by $\bar{u} _{\mu} =\left( -a,0,0,0 \right)$ and $ u_{\mu}^{(\ka)}= \left( -a\phi ^{(\ka)},a\partial _rv^{(\ka)},0,0 \right)$, where $v^{(\ka)}(\eta, r)$ is an arbitrary function.
There are no anisotropic stress in our universe model, so that we have $\phi ^{(\ka)} =\psi ^{(\ka)}$.

The Einstein equations lead to the Friedmann equation for isotropic homogeneous background
\begin{equation}
{\cal H}^2:=\left( \frac{1}{a}\frac{da}{d\eta} \right) ^2={\cal H}_0^2\left(\frac{\Omega_{\rm m}}{a}+\Omega_\Lambda a^2\right),
\label{H-eq}
\end{equation}
where ${\cal H}_0$ is the present value of ${\cal H}$,
\begin{equation}
\Omega_{\rm m}=\frac{8\pi\bar{\rho}_0}{3{\cal H}_0^2}~~~~{\rm and}~~~~
\Omega_\Lambda=\frac{\Lambda}{3{\cal H}_0^2},
\end{equation}
where $\bar{\rho}_0$ is the background energy density at $\eta=\eta_0$. 
The Einstein equations lead to the equations for the linear isotropic perturbations;   
\begin{eqnarray}
\delta ^{(\ka)}=\frac{1}{4\pi a^2 \brho }\left(  -3{\cal H}(\dot{\phi}^{(\ka)}+{\cal H}\phi ^{(\ka)})+\Delta \phi ^{(\ka)} \right),
\label{Eeq_p0}
\end{eqnarray}
\begin{eqnarray}
v^{(\ka )}= -\frac{1}{4\pi a^2\brho}\left( \dot{\phi}^{(\ka)}+{\cal H}\phi ^{(\ka)}\right) ,
\label{Eeq_p1}
\end{eqnarray}
\begin{eqnarray}
\ddot{\phi}^{(\ka)}+3{\cal H}\dot{\phi} ^{(\ka)} +\left( 2\frac{\ddot{a}}{a}-{\cal H}^2\right) \phi ^{(\ka)}=0,
\label{Eeq_p2}
\end{eqnarray}
where a dot denotes a partial differentiation with respect to $\eta$.

The general solution of Eq.~(\ref{Eeq_p2}) is represented by the linear superposition of the growing mode $D_{+}(\eta)$ and the decaying mode $D_{-}(\eta)$, which are defined as 
\begin{equation}
D_{+}(\eta):={\cal H} _0^2\left(\frac{ {\cal H}(a)}{a^2}\int ^{a}_0\frac{1}{{\cal H}^{3}(b)} db \right) ~~~~~{\rm and}~~~~~D_{-}(\eta):=\frac{ {\cal H}(a)}{{\cal H} _0 a^2}.
\label{Dpm-sol}
\end{equation}
Hereafter, we assume that the decaying mode does not exist, since this assumption is consistent with the inflationary universe scenario.  
Accordingly, isotropic inhomogeneities have one functional degree, and
we have 
\begin{eqnarray}
\phi ^{(\ka)}(\eta ,r)=f(r) D_{+}(\eta) ,
\label{delta-f}
\end{eqnarray}
where $f(r)$ is an arbitrary function of the radial coordinate $r$.  In this paper, we paramaetrize $f(r)$ as 
\begin{eqnarray}
f(r)=A\exp \left( -\frac{r^2}{2R^2}\right),
\label{fgauss}
\end{eqnarray}
where $A$ and $R$ are arbitrarily constant.
The value of the isotropic inhomogeneous density perturbation is 0 at $r=\infty$ in our universe model.
We define the Fourier transform of $f(r)$ as
\begin{eqnarray}
\tilde{f}(k)&:=&\int \frac{d^3x}{(2\pi)^{3/2}}e^{-i{\bf k}\cdot{\bf x}}f(r) \cr \cr
&=&AR^3 \exp \left( -\frac{R^2k^2}{2}\right).
\label{fgaussk}
\end{eqnarray}

\section{Gravitational wave background induced by nonlinear interactions between isotropic inhomogeneities and anisotropic inhomogeneous density perturbations}\label{Sec3}

\subsection{Evolution Equations}
To compute the gravitational wave background induced by nonlinear interactions between isotropic inhomogeneities and anisotropic inhomogeneous density perturbations,
we begin with the following perturbed metric 
\begin{eqnarray}
ds^2&=&a^2\left[ -\left(1+2\ka fD_{+} +2 \ep \phi ^{(\ep)} (\eta,{\bf x})\right)d\eta^2 +\left(1-2\ka fD_{+}-2 \ep \psi ^{(\ep)} (\eta,{\bf x}) \right) \left( dr^2+r^2d\Omega ^2\right)  \right. \cr \cr
& &{}\left. + \ka \ep h_{ij}(\eta,{\bf x})dx^i dx ^j \right],
\label{metric}
\end{eqnarray}
where $\ep$ is a positive dimensionless small parameter ($0<\ep \ll 1$).
$\ep$-order  perturbations $\phi ^{(\ep)}$ and $\psi ^{(\ep)}$ are part of unaffected by isotropic inhomogeneities in anisotropic inhomogeneous perturbations, i.e., $\ep$-order perturbations are the same as anisotropic inhomogeneous linear perturbations in the FLRW universe model.
$\ka \ep$-order anisotropic inhomogeneous perturbations $h_{ij}$ are influenced by isotropic inhomogeneities.
Usually, $\ep$-order perturbations have vector and tensor modes and these perturbations produce $\ka \ep$-order tensor modes.
In this paper, we have ignored $\ep$-order vector and tensor modes.

We have assumed that our universe model is filled with only the non-relativistic matter and the cosmological constant, so that we have $\phi ^{(\ep)} =\psi ^{(\ep)}$. 
We assume that $\ep$-order perturbations have only growing mode, so that we have $\phi^{(\ep)}(\eta,{\bf x})=\phi^{(\ep)}(\eta_0, {\bf x})\frac{D_+(\eta)}{D_+(\eta_0)}$.
$\phi^{(\ep)}(\eta_0, {\bf x})$ is characterized by power spectrum $P^{(\ep)}(k)$ which is defined as
\begin{eqnarray}
\langle \tilde{\phi} ^{(\ep)}(\eta _0, {\bf k}) \tilde{\phi} ^{(\ep)}({\eta _0, \bf k}')  \rangle =(2\pi )^3 \delta ^{(3)}({\bf k}-{\bf k}') P^{(\ep)}(k), 
\end{eqnarray}
where 
\begin{eqnarray}
 \tilde{\phi} ^{(\ep)}(\eta, {\bf k}) = \int \frac{d^3x}{(2\pi)^{3/2}}{\rm e}^{-i{\bf k}\cdot {\bf x}}\phi ^{(\ep)} (\eta, {\bf x}).
\end{eqnarray}
$P^{(\ep)}$ can be written as
\begin{eqnarray}
{\cal H}_0^3P^{(\ep)}(k)=A^{(\ep)}\left( \frac{k}{{\cal H}_0}\right) ^{n_s-4}T(k),
\end{eqnarray}
where $A^{(\ep)}$ and $n_s$ are constant and $T(k)$ is the transfer function. 
In this paper, we adopt the fitting formula to calculate the transfer function developed by Eisenstein et al.\cite{Eisenstein:1997jh}.

The $\ka \ep$-order Einstein tensor is
\begin{eqnarray}
G_{ij}^{(\ka \ep)}(\eta, {\bf x})&=&\ka \ep \left[ 
\frac{1}{2}\left( \partial _\eta^2 +2{\cal H} \partial _\eta -\Delta \right) h ^{(\ka \ep)}_{ij}+4D_+\left( \partial _i \phi _{(\ep)} \partial _j f +\partial _i f \partial _j \phi_{(\ep)} \right)  \right. \cr
& &{}+8D_+ \left( \phi_{(\ep)} \partial _i \partial _j f +f \partial _i \partial _j \phi_{(\ep)}\right) \cr
& &{}\left. +({\rm diagonal\  part}) \delta _{ij}\right].
\end{eqnarray}
The $\ka \ep$-order stress-energy tensor of the non-relativistic matter is
\begin{eqnarray}
T_{ij}^{(\ka \ep)}(\eta, {\bf x})=\ka \ep\bar{\rho}( u _{i}^{(\ka)} (\eta, r) u_{j}^{(\ep)} (\eta, {\bf x})+u _{i}^{(\ep)} (\eta, {\bf x}) u_{j}^{(\ka)}(\eta, r)),
\end{eqnarray}
where 
\begin{eqnarray}
u ^{(\ep)}_i(\eta, {\bf x}) = a\partial _i v^{(\ep)}(\eta, {\bf x}),
\end{eqnarray}
\begin{eqnarray}
v^{(\ep)}(\eta, {\bf x})= -\frac{1}{4\pi a^2\brho}\left( \dot{\phi}^{(\ep)} +{\cal H}\phi ^{(\ep)} \right).
\label{v1}
\end{eqnarray}

To derive evolution equations of the gravitational wave background, we act on the spatial components of the Einstein equations with the projection tensor $P _{ijlm}$ which extract the transverse trace-free part of a two-index tensor.
$P _{ijlm}$ is defined through its action on a two-index tensor $F^{lm} (\eta,{\bf x})$
\begin{eqnarray}
P _{ijlm} F^{lm}(\eta,{\bf x}) &:=& \int \frac{d^3k}{(2\pi) ^{3/2}} e_{ij}^{+}({\bf k}) \int \frac{d^3x'}{(2\pi) ^{3/2}} e^{+}_{lm}({\bf k}) e^{i{\bf k}\cdot({\bf x}-\bf{x}')} F^{lm} (\eta,{\bf x}')\cr
& &{}+\int \frac{d^3k}{(2\pi) ^{3/2}} e_{ij}^{\times}({\bf k}) \int \frac{d^3x'}{(2\pi) ^{3/2}} e^{\times}_{lm}({\bf k}) e^{i{\bf k}\cdot({\bf x}-\bf{x}')} F^{lm} (\eta,{\bf x}'),
\label{P}
\end{eqnarray}
where
\begin{eqnarray}
e_{ij}^{+}({\bf k}) =\frac{1}{\sqrt{2}}\left( e_{i}({\bf k}) e_{j}({\bf k}) -\bar{e}_{i}({\bf k})  \bar{e}_{j}({\bf k})  \right),
\label{e}
\end{eqnarray}
and 
\begin{eqnarray}
e_{ij}^{\times}({\bf k}) =\frac{1}{\sqrt{2}}\left( e_{i}({\bf k}) \bar{e}_{j}({\bf k}) +\bar{e}_{i}({\bf k})  e_{j}({\bf k})  \right),
\label{bare}
\end{eqnarray}
where $ e_{i}({\bf k})$ and $\bar{e}_{i}({\bf k})$ are orthonormal basis vectors orthogonal to ${\bf k}$.
$e_i({\bf k})$ and $\bar{e}_i({\bf k})$ leave the rotational degree in the plain orthogonal to ${\bf k}$.
Calculating the transverse trace-free spatial part of the $\ka \ep$-order Einstein equations yields
\begin{eqnarray}
 \ddot{h}_{ij}+ 2{\cal H}\dot{h}_{ij}-\Delta h_{ij}= -2 P _{ijlm}S^{lm},
\label{PG-PT}
\end{eqnarray}
where
\begin{eqnarray}
S_{lm}(\eta,{\bf x})&=&-16\pi a^2 \brho \left( \partial _l v^{(\ep)} \partial _m v^{(\ka)}+\partial _l v^{(\ka)} \partial _m v^{(\ep)} \right) 
+4D_+\left( \partial _l \phi ^{(\ep)} \partial _m f +\partial _l f\partial _m \phi ^{(\ep)} \right) \cr
& &{}+8 D_+\left( \phi ^{(\ep)} \partial _l \partial _m f+f \partial _l \partial _m \phi ^{(\ep)}\right).
\label{Slm}
\end{eqnarray}
Since the equation (\ref{PG-PT}) has source term, the gravitational wave background is induced by nonlinear interactions between isotropic inhomogeneities and anisotropic inhomogeneous density perturbations.
We define the Fourier transform of the gravitational wave background as
\begin{eqnarray}
h_{ij}(\eta, {\bf x}):=\int \frac{d^3k}{(2\pi) ^{3/2}}e^{i{\bf k \cdot x}}\left[ h^{+}(\eta, {\bf k})e_{ij}^{+}({\bf k})+h^{\times}(\eta, {\bf k})e_{ij}^{\times}({\bf k})\right]
\label{hk}
\end{eqnarray}
In Fourier space, evolution equations of the gravitational wave background are
\begin{eqnarray}
 \ddot{h}^{I}+ 2{\cal H}\dot{h}^{I}+k^2 h^{I}=S^{I}(\eta, {\bf k}),
\label{basiceq-1}
\end{eqnarray}
where $I=+, \times$ and
\begin{eqnarray}
S^{I}(\eta,{\bf k})&=&4\int \frac{d^3p}{(2\pi) ^{3/2}} e^{I}_{ij}({\bf k}) p^i p^j 
 \frac{\tilde{\phi}^{(\ep)}(\eta_0, {\bf p})}{D_+(\eta_0)} \tilde{f}({\bf k}-{\bf p}) \cr \cr
 & &\times \left[ \frac{1}{a^2\bar{\rho}\pi} \left( \dot{D}_+ +{\cal H}D_+ \right) ^2+4 D_+^2 \right].
\label{Sk}
\end{eqnarray}

In order to obtain a particular solution of Eq.~(\ref{basiceq-1}), we describe $h^{I}(\eta ,{\bf k})$ as
\begin{eqnarray}
h^{I}(\eta ,{\bf k})=T_{h}(\eta, k)\hat{S}^{I}({\bf k}),
\label{TS}
\end{eqnarray}
where $\hat{S}^{I}({\bf k})$ is
\begin{eqnarray}
\hat{S}^{I}({\bf k})&=&\int \frac{d^3p}{(2\pi) ^{3/2}} e^{I}_{ij}({\bf k}) p^i p^j 
 \frac{\tilde{\phi}_{(\ep)}(\eta_0, {\bf p})}{D_+(\eta_0)} \tilde{f}({\bf k}-{\bf p}).
\end{eqnarray}
$T_{h}(\eta, k)$ represents time evolve of the gravitational wave background and does not depend on spatial variation of isotropic inhomogeneities and anisotropic inhomogeneous density perturbations. 
Substituting Eq.~(\ref{TS}) into Eq.~(\ref{basiceq-1}), and we have
\begin{eqnarray}
\ddot{T}_h +2{\cal H}\dot{T}_h +k^2T_h  =4 \left[ \frac{1}{a^2\bar{\rho}\pi} \left( \dot{D}_+ +HD_+ \right) ^2+4 D_+^2 \right].
\label{Ts}
\end{eqnarray}

Initial conditions of Eq.~(\ref{Ts}) is given as follows.
Since we are interested in the gravitational waves background induced by nonlinear interactions, we assume that there are no gravitational wave background at early universe;
\begin{eqnarray}
T_h(\eta,k)|_{a=0}=0,
\label{boundary-th1}
\end{eqnarray}
Since Eq.~(\ref{Ts}) has regular singular point at $a=0$, we assume following initial conditions to obtain the smooth solution in all $a$;
\begin{eqnarray}
\partial _{\eta}T_h(\eta,k)|_{a=0}=0.
\label{boundary-th2}
\end{eqnarray}

\subsection{Power spectrum and relative energy density}
We calculate two quantities, the power spectrum $P_{h}(\eta, {\bf k},{\bf k}')$ and the relative energy density $\Omega _{\rm GW}(k, \eta)$, to know properties of the gravitational wave background.
We define $P_{h}(\eta, {\bf k},{\bf k}')$ as
\begin{eqnarray}
P_{h}(\eta, {\bf k},{\bf k}')&:=&\langle  \left( h^{+}(\eta,{\bf k})e_{ij}^{+}({\bf k})+h^{\times}(\eta,{\bf k})e_{ij}^{\times}({\bf k}) \right) \cr
& &{} \left( h^{+*}(\eta,{\bf k}')e^{+ij}({\bf k}')+h^{\times *}(\eta,{\bf k}')e^{\times ij}({\bf k}') \right) \rangle.
\label{phk}
\end{eqnarray}
As mentioned above, $e_i$ and $\bar{e}_i$ leave the rotational degree in the plain orthogonal to ${\bf k}$, but $P_{h}$ does not depend on how to fix $e_i$ and $\bar{e}_i$.
$P_{h}$ depends on only three variables, $k:=|{\bf k}|$, $k':=|{\bf k}'|$ and $ \gamma :=\cos ^{-1}\left( \frac{{\bf k}\cdot{\bf k}'}{k k'} \right)$, from spherical symmetry.

We define the relative energy density $\Omega _{\rm GW}(k, \eta)$ as
\begin{eqnarray}
\Omega _{\rm GW}(k, \eta):=\frac{8\pi}{3{\cal H}^2} \frac{d \hat{\rho} _{\rm GW}(k, \eta)}{d \log k},
\label{ogw}
\end{eqnarray}
where 
\begin{eqnarray}
\hat{\rho} _{\rm GW}(k, \eta):=\frac{1}{32\pi} \int _{V_k} \frac{d^3p}{(2\pi )^{3/2}} \int _{V_k} \frac{d^3 p'}{(2\pi )^{3/2}} pp' P_{h}(\eta, {\bf p},{\bf p}'),
\end{eqnarray}
where $V_k$ is a sphere of radius $k$.
$d\hat{\rho} _{\rm GW}(k, \eta)$ means the energy density of the gravitational wave background contained in the wave number range $k$ to $k + dk$.
If $P_{h}$ can be describe as 
\begin{eqnarray}
P_{h}(\eta, {\bf k},{\bf k}')= 2\pi ^2 \delta ^{(3)}({\bf k}-{\bf k}') \frac{P_{h}^{(0)}(\eta, k)}{k^3},
\label{Ph0}
\end{eqnarray}
we have
\begin{eqnarray}
\Omega _{\rm GW}(k, \eta)=\frac{1}{12{\cal H}^2} k^2P_{h}^{(0)}(\eta, k),
\label{ogw0}
\end{eqnarray}
where $P_{h}^{(0)}$ is arbitrary function of $\eta$ and $k$.
The primordial inflationary gravitational wave background in the FLRW universe model is one example that $P_{h}(\eta, {\bf k},{\bf k}')$ can be described as Eq.~(\ref{Ph0}).
Eq.~(\ref{ogw0}) means that, if we calculate the relative energy density of the primordial inflationary gravitational wave background in the FLRW universe model, the relative energy density which we define in Eq.~(\ref{ogw}) coincide with that of used in studies of the primordial inflationary gravitational wave background in the FLRW universe model.
The energy density of the gravitational wave background depends on the background of perturbations.
In our universe model, we assume that the background of the gravitational wave background is the background FLRW universe model.

\section{Numerical result}\label{Sec4}
Before performing numerical integral of Eq.~(\ref{basiceq-1}), we choose the parameters in the background FLRW universe and $\ep$-order perturbations consistent with Planck results\cite{Ade:2015xua}, $\Omega _{\rm \Lambda}=0.6911$, $\Omega _{\rm b}=0.0486$, ${\cal H} _0=67.74 {\rm km}{\rm s}^{-1}{\rm Mpc}^{-1}$, $A^{(\ep)}=1.09\times 10^{-8}$ and $n_{\rm s}=0.9667$.

\begin{figure}[h]
 \begin{center}
\includegraphics[width=100mm]{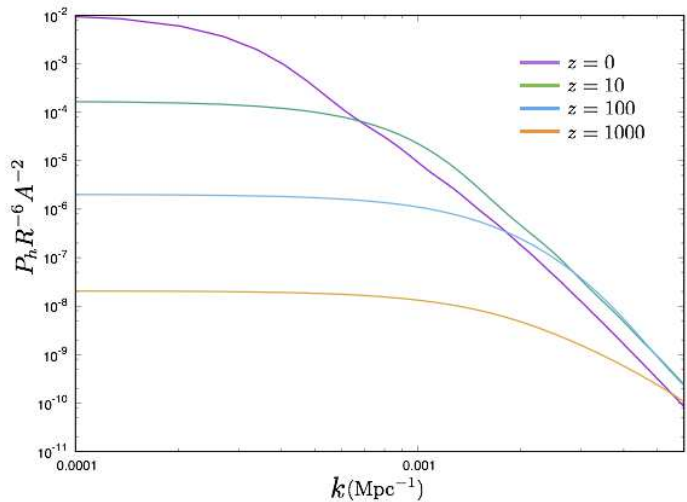}
 \end{center}
 \caption{We depict $P_{h}$ as a function of $k$ in the case of ${\bf k}={\bf k}'$, $R=1{\rm Gpc}$ and with various $z$.}
 \label{fig:1}
\end{figure}

\begin{figure}[h]
 \begin{center}
\includegraphics[width=100mm]{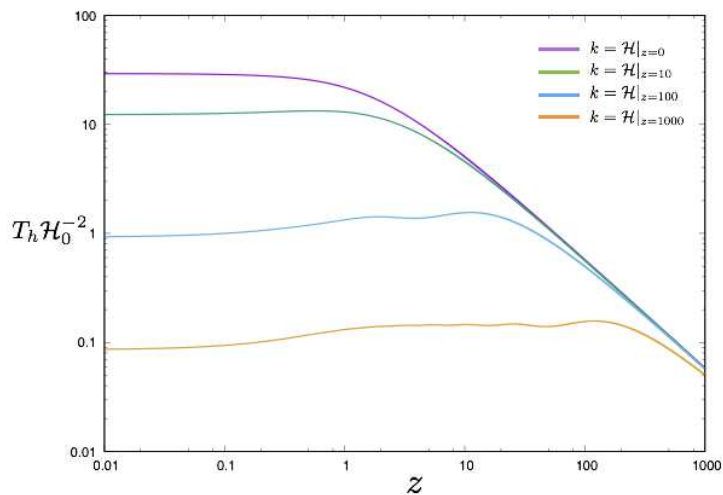}
 \end{center}
 \caption{We depict $T_h$ as a function of $z$ with various $k$.}
 \label{fig:2}
\end{figure}

First, we examine the temporal evolution of the gravitational wave background.
The temporal evolution of $P_{h}$ depend on only $T_{h}$ from Eq.~(\ref{TS}).
In Fig.~\ref{fig:1}, we depict $P_{h}$ as a function of $k$ in the case of ${\bf k}={\bf k}'$, $R=1{\rm Gpc}$ and with various redshift $z$.
In Fig.~\ref{fig:2}, we depict $T_{h}$ as a function of $z$ with various $k$.
It can be seen from Fig.~\ref{fig:2} that $T_{h}$ with $k={\cal H}|_{z=z_c}$ is growing in the domain $z>z_c$ where $\frac{1}{k}$ is larger than horizon scale and almost constant in the domain $z<z_c$ where $\frac{1}{k}$ is smaller than horizon scale, where $z_c=10, 100, 1000$.
This result can be understood from Eq.~(\ref{Ts}).
In the domain $\frac{1}{k}>\frac{1}{{\cal H}}$, we can ignore the oscillation term and the gravitational wave background grows slowly due to production by the source term and friction caused by cosmic expansion.
In the domain $\frac{1}{k}<\frac{1}{{\cal H}}$, we can ignore the friction term and $T_{h}$ oscillate about the source term.
In the case $k={\cal H}|_{z=0}$, $T_{h}$ hardly grows in the domain $z<1$.
This is because cosmological constant become important in the domain $z<1$ and growth of the gravitational wave background is hindered by the accelerated expansion.

\begin{figure}[h]
 \begin{center}
\includegraphics[width=100mm]{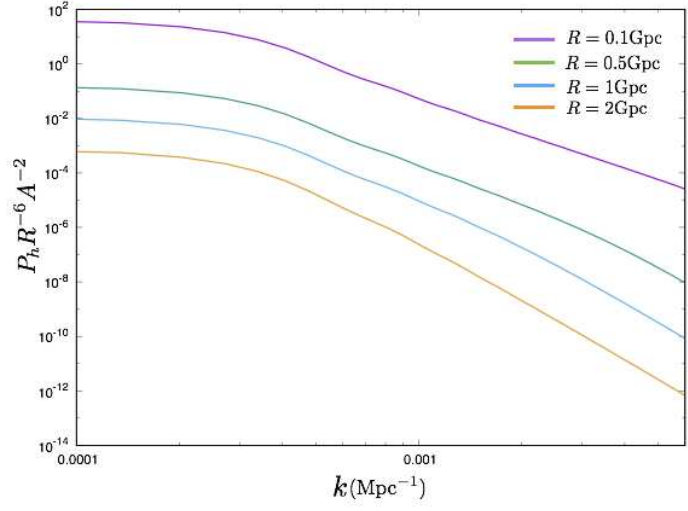}
 \end{center}
 \caption{We depict $P_{h}$ as a function of $k$ in the case of ${\bf k}={\bf k}'$, $z=0$ and with various $R$.}
 \label{fig:3}
\end{figure}

\begin{figure}[h]
 \begin{center}
\includegraphics[width=100mm]{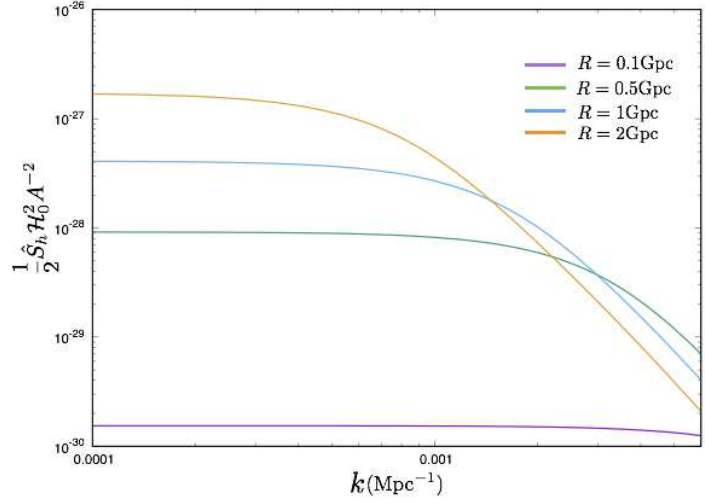}
 \end{center}
 \caption{We depict the $\hat{S}_h$ as a function of $k$ in the case of ${\bf k}={\bf k}'$ and with various $R$.}
 \label{fig:4}
\end{figure}

We examine $R$-dependence of the gravitational wave background.
Substituting Eq.~(\ref{TS}) into Eq.~(\ref{phk}) and we have 
\begin{eqnarray}
P_{h}(\eta, {\bf k},{\bf k}')=T_h(k)T_h(k') \hat{S}_h({\bf  k},{\bf k}'),
\end{eqnarray}
where
\begin{eqnarray}
\hat{S}_h({\bf  k},{\bf k}')&:=&
\langle  \left( \hat{S}^{+}({\bf k})e_{ij}^{+}({\bf k})+\hat{S}^{\times}({\bf k})e_{ij}^{\times}({\bf k}) \right) \cr
& &{} \left( \hat{S}^{+*}({\bf k}')e^{+ij}({\bf k}')+\hat{S}^{\times *}({\bf k}')e^{\times ij}({\bf k}') \right) \rangle.
\end{eqnarray}
Thus, in order to know the $R$-dependence of $P_{h}$, we have to pay attention to only  $\hat{S}_h$, since $T_{h}$ does not depend on $R$.
In Fig.~\ref{fig:3}, we depict $P_{h}$ as a function of $k$ in the case of ${\bf k}={\bf k}'$, $z=0$ and with various $R$.
In Fig.~\ref{fig:4}, we depict $\hat{S}_h$ as a function of $k$ in the case of ${\bf k}={\bf k}'$ and with various $R$.
It can be seen from Fig.~\ref{fig:4} that the amplitude of gravitational wave background become small in the domain $\frac{1}{k}<R$.
This is because the gravitational wave background is produced by distorting isotropic inhomogeneities due to nonlinear interactions between isotropic inhomogeneities and anisotropic inhomogeneous density perturbations.

\begin{figure}[h]
 \begin{center}
\includegraphics[width=100mm]{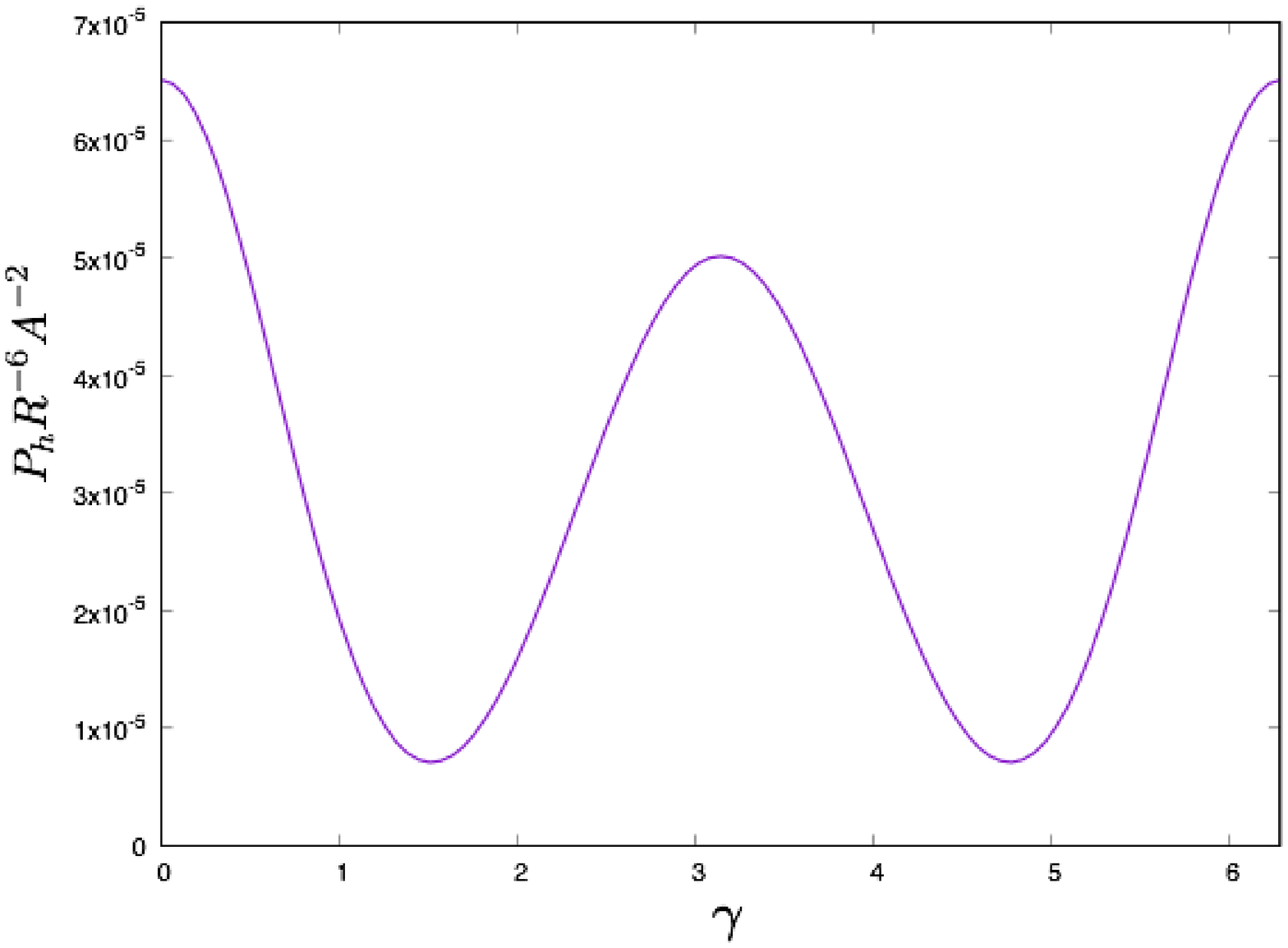}
 \end{center}
 \caption{ We depict the $P_{h}$ as a function of $\gamma$ in the case of $k=k'=3{\cal H}_0$, $R=1{\rm Gpc}$ and $z=0$.}
 \label{fig:5}
\end{figure}

In Fig.~\ref{fig:5}, we depict $P_{h}$ as a function of $\gamma$ in the case of $k=k'=3{\cal H}_0$, $R=1{\rm Gpc}$ and $z=0$.
$P_{h}$ is symmetric about the $\gamma=\pi$ from spherical symmetry.

\begin{figure}[h]
 \begin{center}
\includegraphics[width=100mm]{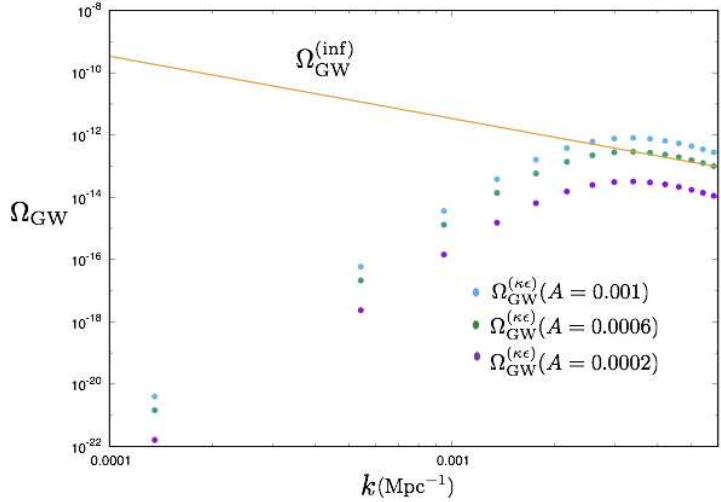}
 \end{center}
 \caption{We depict $\Omega _{\rm GW}$ of the gravitational wave background induced by nonlinear interactions and the primordial inflationary gravitational wave background as a function of $k$ in the case of $z=1100$, $R=1{\rm Gpc}$ and with various $A$.}
 \label{fig:6}
\end{figure}

\begin{figure}[h]
 \begin{center}
\includegraphics[width=100mm]{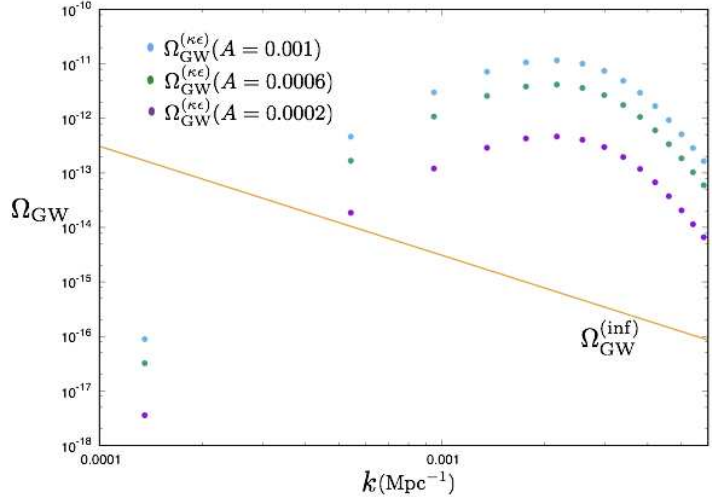}
 \end{center}
 \caption{We depict the same as Fig.~\ref{fig:6} but in the case of $z=0$, $R=1{\rm Gpc}$ and with various $A$.}
 \label{fig:7}
\end{figure}

We compare the relative energy density of the gravitational wave background induced by nonlinear interactions between isotropic inhomogeneities and anisotropic inhomogeneous density perturbations with the primordial inflationary gravitational wave background to discuss observability.
The relative energy density of the primordial inflationary gravitational wave background $\Omega _{\rm GW}^{({\rm inf})}$ can be expressed as \cite{Baumann:2007zm}
\begin{eqnarray}
\Omega _{\rm GW}^{({\rm inf})}(k, \eta)=A_{\rm GW}r_0 \Delta ^2 _{R}(k_0)\left( \frac{k}{k_0}\right) ^{n_t}\left\{ \begin{array}{ll}
\frac{a_{\rm eq}}{a} \left( \frac{k}{k_{\rm eq}}\right) ^{-2}~~~~(k<k_{\rm eq}) \\
\frac{a_{\rm eq}}{a}  ~~~~\hspace{1.55cm} (k>k_{\rm eq}) \\
\end{array} \right.
\end{eqnarray}
where $A_{\rm GW}=4.2\times 10^{-2}$, $k_0=0.002 {\rm Mpc}^{-1}$, $a_{\rm eq}=\frac{1}{3400}$, $k_{\rm eq}=0.01{\rm Mpc}^{-1}$, $\Delta ^2 _{R}(k_0)=2.5\times 10^{-9}$, $r_0$ is the tensor-to-scalar ratio evaluated on $k_0$ and $n_t$ is constant.
We choose the parameters which characterize the primordial inflationary gravitational wave background as $r_0=0.001$ and $n_t=0$, since LiteBIRD promise to determine $r_0$ with a precision of $\delta r_0 = \mathcal{O}(10^{-3})$.
We depict the relative energy density of the gravitational wave background induced by nonlinear interactions $\Omega _{\rm GW}^{(\ka \ep)}$ and $\Omega _{\rm GW}^{({\rm inf})}$ as a function of $k$ in the case of $z=1100$, $R=1{\rm Gpc}$ and with various $A$ in Fig~\ref{fig:6} and  $z=0$, $R=1{\rm Gpc}$ and with various $A$ in Fig~\ref{fig:7}.
It can be seen from Fig.~\ref{fig:6} that, near the scale $k = {\cal H}|_{z=1100}\approx 0.004$, the relative energy density of the gravitational wave background induced by nonlinear interactions with $A=0.0006$ is almost the same value as that of the primordial inflationary gravitational wave background.
$\delta ^{(\ka)}$ at the center and present time is about $0.06$, if $R=1{\rm Gpc}$ and $A=0.0006$.
Thus, the gravitational wave background induced by nonlinear interactions may have influence on the CMB as much as the primordial inflationary gravitational wave background whose tensor-to-scalar ratio is 0.001, if  $6\%$-level isotropic deviations from homogeneity exist.
It can be seen from Fig.~\ref{fig:7} that as these gravitational waves time development, $\Omega _{\rm GW}^{(\ka \ep)}$ becomes larger than $\Omega _{\rm GW}^{({\rm inf})}$ on large scales.
This is because the gravitational wave background induced by nonlinear interactions is growing by time development, in contrast the primordial inflationary gravitational wave background redshift on all scales.

\section{Summary and discussion}\label{Sec5}
We studied the gravitational wave background induced by nonlinear interactions between isotropic inhomogeneities and anisotropic inhomogeneous density perturbations.
We assumed that the amplitude of isotropic inhomogeneities is small and, on large scales, we described the universe as the FLRW universe model with isotropic inhomogeneous perturbation.
Then, we solved perturbation equations up to the order that isotropic inhomogeneities and anisotropic inhomogeneous density perturbations are coupled to obtain the gravitational wave background induced by nonlinear interactions between isotropic inhomogeneities and anisotropic inhomogeneous density perturbations.
Our results are Fig.~\ref{fig:1}--Fig.~\ref{fig:7}.
Unlike the primordial inflationary gravitational wave background in the FLRW universe model, the gravitational wave background induced by nonlinear interactions grows as time evolves.
This is because, there are source terms in the evolution equation of this gravitational wave background.
The gravitational wave background induced by nonlinear interactions is hardly produced in wavelengths that is smaller than the scale of isotropic inhomogeneities, since the gravitational wave background is produced by distorting isotropic inhomogeneities due to nonlinear interactions.

Our result implies that we can observe the gravitational wave background induced by nonlinear interactions between isotropic inhomogeneities and anisotropic inhomogeneous density perturbations in the future CMB observation project, such as LiteBIRD that promise to determine $r_0$ with a precision of $\delta r_0 = \mathcal{O}(10^{-3})$, if there are isotropic inhomogeneities whose amplitude of density perturbations is about 0.06 in the universe.
In order to know the influence of this gravitational wave background on the CMB in more detail, we need to solve the Boltzmann equation.
This is our future work.

\section*{Acknowledgments}
We are grateful to Ken-ichi Nakao, Hideki Ishihara, Ryusuke Nishikawa and colleagues in the group of elementary 
particle physics and gravity at Osaka City University for useful discussions and helpful comments.


\begin{thebibliography}{99}
\bibitem{Redlich:2014gga} 
  M.~Redlich, K.~Bolejko, S.~Meyer, G.~F.~Lewis and M.~Bartelmann,
  ``Probing spatial homogeneity with LTB models: a detailed discussion,''
  Astron.\ Astrophys.\  {\bf 570}, A63 (2014)
  [arXiv:1408.1872 [astro-ph.CO]].
  
  
\bibitem{Negishi:2015oga} 
  H.~Negishi, K.~i.~Nakao, C.~M.~Yoo and R.~Nishikawa,
  ``Systematic error due to isotropic inhomogeneities,''
  Phys.\ Rev.\ D {\bf 92}, no. 10, 103003 (2015)
  [arXiv:1505.02472 [astro-ph.CO]].


\bibitem{Clarkson:2009sc} 
  C.~Clarkson, T.~Clifton and S.~February,
  ``Perturbation Theory in Lema\^{\i}tre-Tolman-Bondi Cosmology,''
  JCAP {\bf 0906}, 025 (2009)
  [arXiv:0903.5040 [astro-ph.CO]].


\bibitem{Zibin:2008vj} 
  J.~P.~Zibin,
  ``Scalar Perturbations on Lema\^{\i}tre-Tolman-Bondi Spacetimes,''
  Phys.\ Rev.\ D {\bf 78}, 043504 (2008)
  [arXiv:0804.1787 [astro-ph]].


\bibitem{Nishikawa:2012we} 
  R.~Nishikawa, C.~M.~Yoo and K.~i.~Nakao,
  ``Evolution of density perturbations in large void universe,''
  Phys.\ Rev.\ D {\bf 85}, 103511 (2012)
  [arXiv:1202.1582 [astro-ph.CO]].
  
  
\bibitem{Nishikawa:2013rna} 
  R.~Nishikawa, C.~M.~Yoo and K.~i.~Nakao,
  ``Two-point correlation function of density perturbations in a large void universe,''
  Phys.\ Rev.\ D {\bf 88}, no. 12, 123520 (2013)
  [arXiv:1306.5131 [astro-ph.CO]].
  
  
\bibitem{Nishikawa:2014sga} 
  R.~Nishikawa, K.~i.~Nakao and C.~M.~Yoo,
  ``Comparison of two approximation schemes for solving perturbations in a Lema\^{\i}tre-Tolman-Bondi cosmological model,''
  Phys.\ Rev.\ D {\bf 90}, no. 10, 107301 (2014)
  [arXiv:1407.4899 [astro-ph.CO]].

\bibitem{February:2013qza} 
  S.~February, J.~Larena, C.~Clarkson and D.~Pollney,
  ``Evolution of linear perturbations in spherically symmetric dust spacetimes,''
  Class.\ Quant.\ Grav.\  {\bf 31}, 175008 (2014)
  [arXiv:1311.5241 [astro-ph.CO]].

\bibitem{Meyer:2014qla} 
  S.~Meyer, M.~Redlich and M.~Bartelmann,
  ``Evolution of linear perturbations in Lema\^{\i}tre-Tolman-Bondi void models,''
  JCAP {\bf 1503}, no. 03, 053 (2015)
  [arXiv:1412.3012 [astro-ph.CO]].



\bibitem{Eisenstein:1997jh} 
  D.~J.~Eisenstein and W.~Hu,
  ``Power spectra for cold dark matter and its variants,''
  Astrophys.\ J.\  {\bf 511}, 5 (1997)
  [astro-ph/9710252].
  
  
  
\bibitem{Ade:2015xua} 
  P.~A.~R.~Ade {\it et al.} [Planck Collaboration],
  ``Planck 2015 results. XIII. Cosmological parameters,''
  Astron.\ Astrophys.\  {\bf 594}, A13 (2016)
  [arXiv:1502.01589 [astro-ph.CO]].
  
  
\bibitem{Baumann:2007zm} 
  D.~Baumann, P.~J.~Steinhardt, K.~Takahashi and K.~Ichiki,
  ``Gravitational Wave Spectrum Induced by Primordial Scalar Perturbations,''
  Phys.\ Rev.\ D {\bf 76}, 084019 (2007)
  [hep-th/0703290].

\end{thebibliography}
\end{document}